\documentclass[aps, prb, reprint]{revtex4-2} 
\usepackage{fix-cm}
\usepackage[T1]{fontenc}

\usepackage{amsmath}  
\usepackage{amsfonts} 
\usepackage{graphicx} 
\usepackage{dcolumn}
\usepackage{bm}
\usepackage{xcolor}
\usepackage{float}

\usepackage{titlesec}
\titlespacing*{\section}{0pt}{\baselineskip}{\baselineskip}
\usepackage[utf8]{inputenc}
\usepackage[T1]{fontenc}
\usepackage{mathptmx}
\usepackage{etoolbox}

\begin{document}


\title{Demonstrating Dynamic Stability in Paul Traps: Exploring\\Rotating Saddles with Liquid Nitrogen Droplets}


\author{Laurel Barnett}
\email{ljbarnett@college.harvard.edu}
\author{Aidan Carey}
\email{aidanbcarey@gmail.com}
\author{Robert Hart}
\email{roberthart56@gmail.com}
\author{Daniel Davis}
\email{ddavis@fas.harvard.edu}
\author{Anna Klales}
\email{aklales@g.harvard.edu}
\author{Louis Deslauriers}
\email{louis@physics.harvard.edu}
\affiliation{Department of Physics, Harvard University, Cambridge, MA 02138}


\date{\today}

\begin{abstract}

Rotating saddle potentials provide a compelling visual demonstration of dynamic stability, widely used in undergraduate physics as mechanical analogs to the RF Paul trap. Traditional demonstrations typically rely on rolling ball bearings, whose frictional effects and internal rotation obscure fundamental particle dynamics. We introduce a simple yet significant improvement by employing droplets of liquid nitrogen (LN$_2$), which levitate via the Leidenfrost effect, eliminating rolling dynamics and greatly reducing friction. LN$_2$ droplets clearly illustrate the rotating ponderomotive-like force, producing trajectories closely consistent with theoretical predictions. Using experimental data, we compare the stability threshold and particle trajectories of LN$_2$ droplets and traditional ball bearings. LN$_2$ droplets exhibit a sharply defined and visually distinct stability threshold, transitioning abruptly from unstable to stable motion at a critical rotation frequency. In contrast, ball bearings demonstrate a more gradual threshold, accompanied by trajectories complicated by friction-induced deviations. We present detailed measurements of particle lifetimes and trajectories as functions of dimensionless stability parameters for both symmetric and intentionally asymmetric saddles. These improvements significantly enhance visual and conceptual clarity, reduce common misconceptions related to frictional dynamics, and provide natural opportunities for exploring related phenomena such as the Leidenfrost effect. We also offer practical guidance on assembling and implementing this enhanced demonstration for effective classroom and laboratory instruction.

\end{abstract}

\maketitle 

\section{INTRODUCTION}\label{sec:intro} 

A saddle-shaped surface rotating about its inflection point can dynamically stabilize and confine a particle near its center, a visually compelling demonstration frequently used in undergraduate physics education. This stabilization occurs when the angular velocity of rotation surpasses a critical threshold, converting an otherwise unstable equilibrium into a dynamically stable one \cite{Paul_1990, Rueckner_1995}. Rotating saddle demonstrations vividly illustrate dynamic stability, a fundamental phenomenon central to many physical systems, including RF Paul traps, Rydberg atoms confined by rotating electric fields, Lagrange points of celestial mechanics, and recently identified gravitational saddle potentials near charged binary black hole systems \cite{Kapitza_1951, Whymark_1975, Raab_1987, Bialynicki-Birula_1994, Murray_1999, Dehmelt_1967, Leibfried_2003, Abe_2010, Kurmus_2025}. These examples underscore the broad relevance of rotating saddle dynamics across classical, quantum, and astrophysical contexts.

Typical classroom implementations use spherical objects — often steel ball bearings — whose rolling friction and internal rotation cause their motion to differ significantly from the idealized dynamics of a point particle trapped in a rotating saddle \cite{Thompson_2002, Kirillov_2016, Fan_2017, Borisov_2018}. In practice, observers readily notice these complications: rolling balls often exhibit erratic, swirling trajectories that clearly reflect significant frictional interaction with the rotating saddle. These frictional effects make it difficult to observe cleanly the characteristic dynamics expected from a rotating potential. Moreover, successful demonstrations typically require carefully dropping the ball nearly at rest and precisely at the saddle’s center to maximize trapping lifetimes, a challenging task that further limits the effectiveness of the demonstration.

We introduce a straightforward yet highly effective improvement: replacing rolling spheres with droplets of liquid nitrogen (LN$_2$). Due to the Leidenfrost effect, these droplets levitate on a thin vapor cushion formed by rapid evaporation, eliminating rolling dynamics and greatly reducing friction \cite{Biance_2003, Quere_2013}. Consequently, their motion closely approximates frictionless particle dynamics, substantially enhancing pedagogical clarity.

The primary advantage of LN$_2$ droplets is that their motion clearly reveals the rotating nature of the confining force, producing trajectories consistent with theoretical predictions of a rotating saddle potential. In contrast, rolling balls exhibit trajectories visibly distorted by frictional coupling to the saddle surface, substantially diminishing the instructional clarity of the demonstration. Furthermore, the low-friction LN$_2$ droplets are relatively insensitive to initial positioning and velocity. Unlike rolling balls, LN$_2$ droplets do not require precise placement near the saddle center or careful initial velocity control. As we discuss in this paper, LN$_2$ droplets typically exhibit stable trapping even when initially deposited off-center or in motion.

An additional key benefit is the sharp transition between stability and instability exhibited by LN$_2$ droplets as the rotation frequency crosses the stability threshold. Even minute frequency adjustments produce a distinct, visual shift from stable trapping to rapid escape, making this demonstration particularly effective in highlighting the concept of dynamic stability thresholds.

Real-world analogs to rotating saddle traps, such as RF Paul traps used for precision quantum manipulation, often incorporate slight potential asymmetries to lift degeneracies in particle motion, facilitating laser cooling and quantum control \cite{Deslauriers_2004, Leibfried_2003, Chou_2010, Ladd_2010}. Motivated by this practical relevance, our demonstrations include saddles with both intentional asymmetry and symmetric geometry, allowing clear investigation into how subtle variations in symmetry affect particle trajectories and trapping stability \cite{carey_2025}. 

In this paper, we describe the theoretical background, experimental implementation, and practical classroom use of the LN$_2$ rotating saddle demonstration. We quantify the substantial improvement in visual and quantitative fidelity compared to traditional rolling spheres by mapping the lifetime of LN$_2$ droplets as a function of the stability parameters $a$ and $q$, which we vary experimentally by adjusting the rotation frequency \cite{carey_2025}. We present lifetime data illustrating sharp threshold behavior, representative particle trajectories, and practical details about fabricating the saddles, assembling the demonstration apparatus, and effectively performing this compelling demonstration in educational settings. Beyond enhancing the understanding of dynamic stability and rotating saddle potentials, this demonstration also naturally introduces opportunities to explore related phenomena such as the Leidenfrost effect itself.

\section{BACKGROUND THEORY}\label{sec:rf_vs_rotating} 

The motion of a particle on a rotating saddle surface can be described by considering the gravitational potential in a rotating reference frame. Following the derivation presented in detail by Carey et al.~\cite{carey_2025}, we summarize the key equations and parameters relevant to understanding dynamic stability in our demonstration.

In the rotating frame, the gravitational potential for a saddle-shaped surface can be approximated by the hyperbolic form:

\begin{equation}
U(x', y') = \frac{m g h_0}{r_0^2}\left(\beta x'^2 - y'^2\right),
\end{equation}

where $m$ is the mass of the particle, $g$ is gravitational acceleration, $h_0$ and $r_0$ represent geometric parameters of the saddle (the saddle height and radius, respectively), and $\beta$ characterizes the curvature asymmetry of the saddle. A symmetric saddle corresponds to $\beta = 1$.

Transforming into the laboratory frame rotating with angular frequency $\Omega$, and introducing the dimensionless time $\tau = \Omega t$, the equations of motion take the standard Mathieu-like form~\cite{carey_2025, Thompson_2002}:

\begin{align}
\frac{d^2 x}{d\tau^2} + (a + 2q \cos(2\tau)) x + 2q \sin(2\tau) y &= 0, \\
\frac{d^2 y}{d\tau^2} + (a - 2q \cos(2\tau)) y + 2q \sin(2\tau) x &= 0,
\end{align}

where the dimensionless parameters $a$ and $q$ are given by:
\begin{equation}
a = \frac{g h_0(\beta - 1)}{r_0^2 \Omega^2}, \quad q = \frac{g h_0(\beta + 1)}{r_0^2 \Omega^2}.
\label{eq:a_q_definition}
\end{equation}

These parameters determine the stability of particle trajectories, analogous to the stability regions found in RF Paul traps \cite{Wineland_1998, Deslauriers_2007}. Stability occurs when particle oscillations remain bounded within the saddle’s confines; instability corresponds to rapidly diverging trajectories. Carey et al.~\cite{carey_2025} provide detailed derivations and stability diagrams illustrating how stability conditions depend on these parameters.

Previous experimental implementations have typically employed spherical objects — such as steel balls or plastic spheres — as analog particles. However, rolling motion introduces frictional forces and internal rotation that lead to significant deviations from the idealized point-mass dynamics assumed in theoretical models. Notably, Kirillov and Levi~\cite{Kirillov_2016} showed that rolling friction induces a precession analogous to that observed in a Foucault pendulum, while Fan et al.~\cite{Fan_2017} demonstrated that certain deviations persist even as the ball radius tends to zero, indicating that the breakdown in correspondence with ideal trajectories is not merely a finite-size effect. These non-ideal behaviors obscure the essential features of dynamic stability and motivate the use of friction-minimizing alternatives.

In our experiments, we use liquid nitrogen (LN$_2$) droplets, which levitate via the Leidenfrost effect and closely approximate frictionless point-particle motion. Varying the saddle’s rotation frequency $\Omega$ tunes the dimensionless parameters $a$ and $q$, allowing us to systematically explore the transition between stable and unstable regimes. The equations presented above provide a quantitative framework for interpreting the measured particle lifetimes and trajectories described in the following sections.

\begin{figure}[t]
    \centering
    \includegraphics[width=0.32\textwidth, angle=0,]{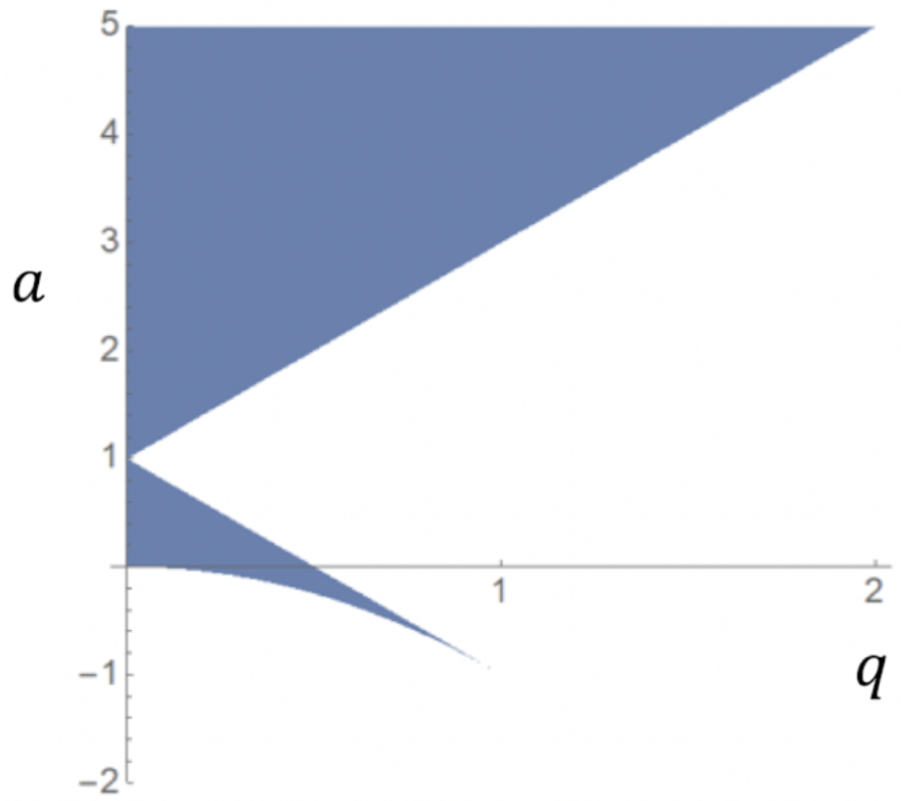}
    \caption{Stability diagram for the rotating saddle potential, showing regions in the $(a, q)$ parameter space where particle motion is stable (shaded blue). The stability boundary is analytically determined and forms a simple triangular region for positive $a$, with a sharp instability at $a = 1$. For negative $a$, stability terminates at $q = 1$, $a = -1$.}
    \label{fig:stability_region}
\end{figure}

\section{EXPERIMENTAL SETUP}

\begin{figure}[h]
    \centering
    \includegraphics[width=90mm]{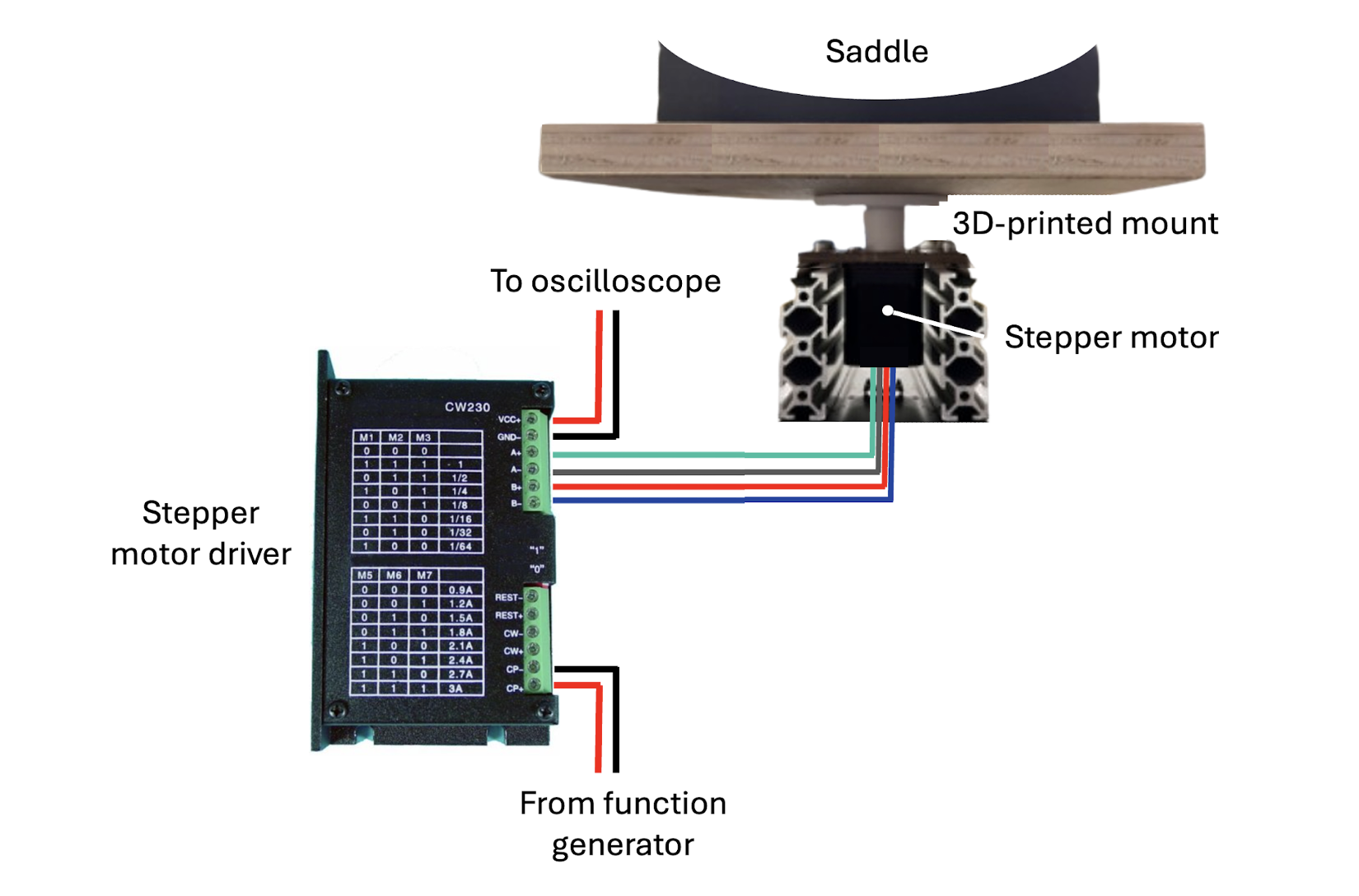}
    \caption{Schematic of the rotating saddle demonstration setup, showing the motor, driver, function generator, and 3D-printed saddle mount.}
    \label{fig:saddle_schematic}
\end{figure}

The core of the demonstration consists of a 3D-printed saddle mounted on a motor via a custom 3D-printed adapter, as shown in Fig.~\ref{fig:saddle_schematic}.

The saddle is rotated by a stepper motor controlled by a microstep driver, which receives a frequency signal from a function generator. The driver is powered by an external voltage source. The output frequency of the function generator is monitored via a BNC connection to an oscilloscope, allowing precise control and real-time feedback on the rotation speed.

\section{DEMONSTRATION PROCEDURE}
\label{sec:demo_procedure}

For large lecture settings, the demonstration is best projected using an overhead camera with a frame rate of at least 60 frames per second. When using LN$_2$ droplets, additional lighting (e.g., a directional lamp angled from above) can greatly enhance visibility. Further practical considerations are described below.

In our demonstrations, we use two 3D-printed saddles with different asymmetry coefficients: a nearly symmetric saddle with $\beta = 1.06$, and an asymmetric saddle with $\beta = 2.67$. Both saddles share identical geometric dimensions, characterized by a saddle height $h_0 = 2.5\,\text{cm}$ and a radius $r_0 = 9\,\text{cm}$. These parameters define the saddle curvature and directly influence the particle dynamics. The nearly symmetric saddle ($\beta=1.06$) corresponds to a vertical deviation of approximately 1.5\,mm at the saddle's edge ($r_0 = 9\,\text{cm}$) compared to a perfectly symmetric saddle ($\beta = 1.06$). Comparing these two saddles illustrates how small differences in geometry significantly affect the sharpness of the dynamic stability threshold and resulting particle trajectories. Unless otherwise stated, rotation frequencies and specific threshold behaviors described in this section refer to the asymmetric saddle ($\beta = 2.67$).

\subsection{Trapping liquid nitrogen droplets}
\label{sec:ln2_demo}

Each LN$_2$ bead is only a few millimeters in diameter, so for classroom visibility, the camera and lighting setup described above is particularly important.

\begin{figure}[h]
\centering
\includegraphics[width=88mm]{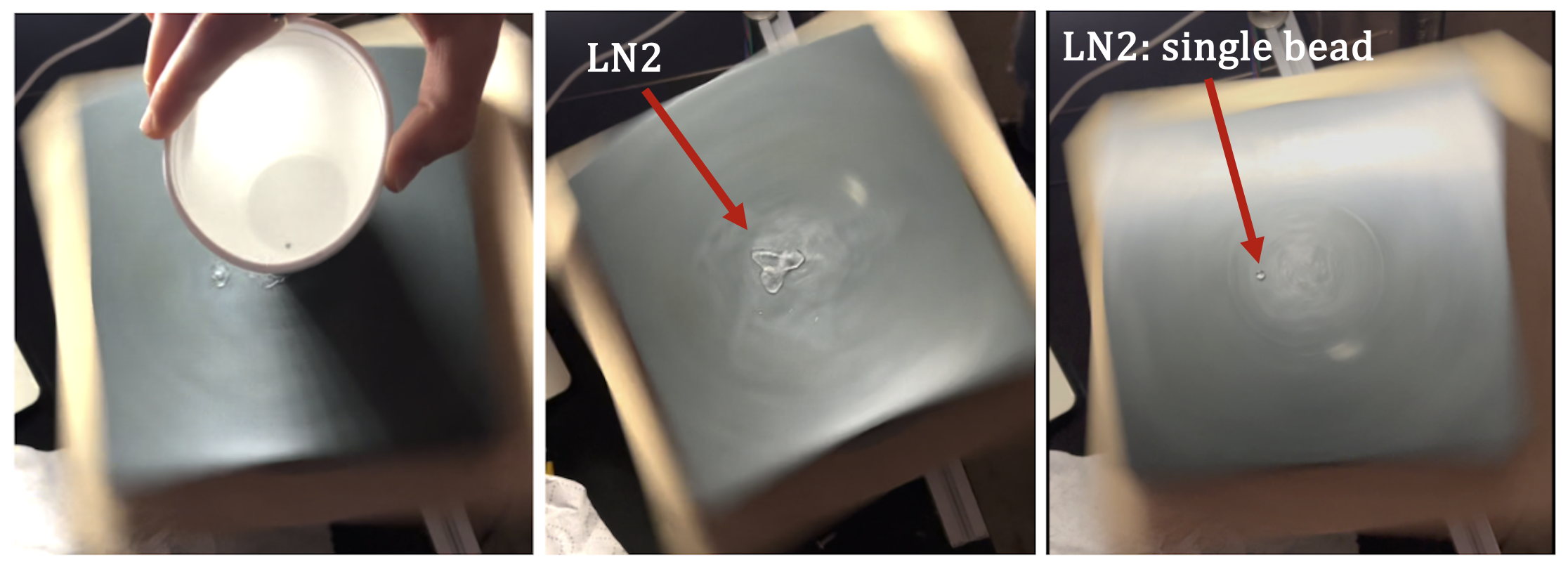}
\caption{Evolution of liquid nitrogen droplets on a rotating saddle over time.}
\label{fig:ln2_evolution}
\end{figure}

Begin the demonstration by presenting students with a stationary saddle. Optionally, instructors may first display a photograph or CAD rendering of the saddle's profile to illustrate its geometry. Dispense a small quantity of LN$_2$ at the center of the saddle; students will immediately observe the liquid nitrogen sliding down the negative curvature of the saddle. Fig.~\ref{fig:ln2_evolution} provides a visual example of how LN$_2$ droplets evolve on the rotating saddle after dispensation; further details on optimal dispensing procedures and considerations are discussed later in this section. Before proceeding, instructors can engage students by prompting them to predict factors that might enable the LN$_2$ bead to remain trapped near the saddle’s center once rotation begins. This approach aligns with research-based best practices for interactive demonstrations, which emphasize prediction, discussion, and conceptual reasoning to enhance student understanding~\cite{Miller_2014, Mazur_1997}.

Next, turn on the stepper motor at a rotational frequency below the expected stability threshold. For the saddle with positive asymmetry ($\beta = 2.67$), start at or below approximately 1.15 rotations per second (rps). After student predictions, dispense LN$_2$ onto the rotating saddle multiple times to clearly demonstrate instability—each bead will rapidly slide off the saddle surface.

\begin{figure}[h]
  \centering
  \includegraphics[width=\columnwidth]{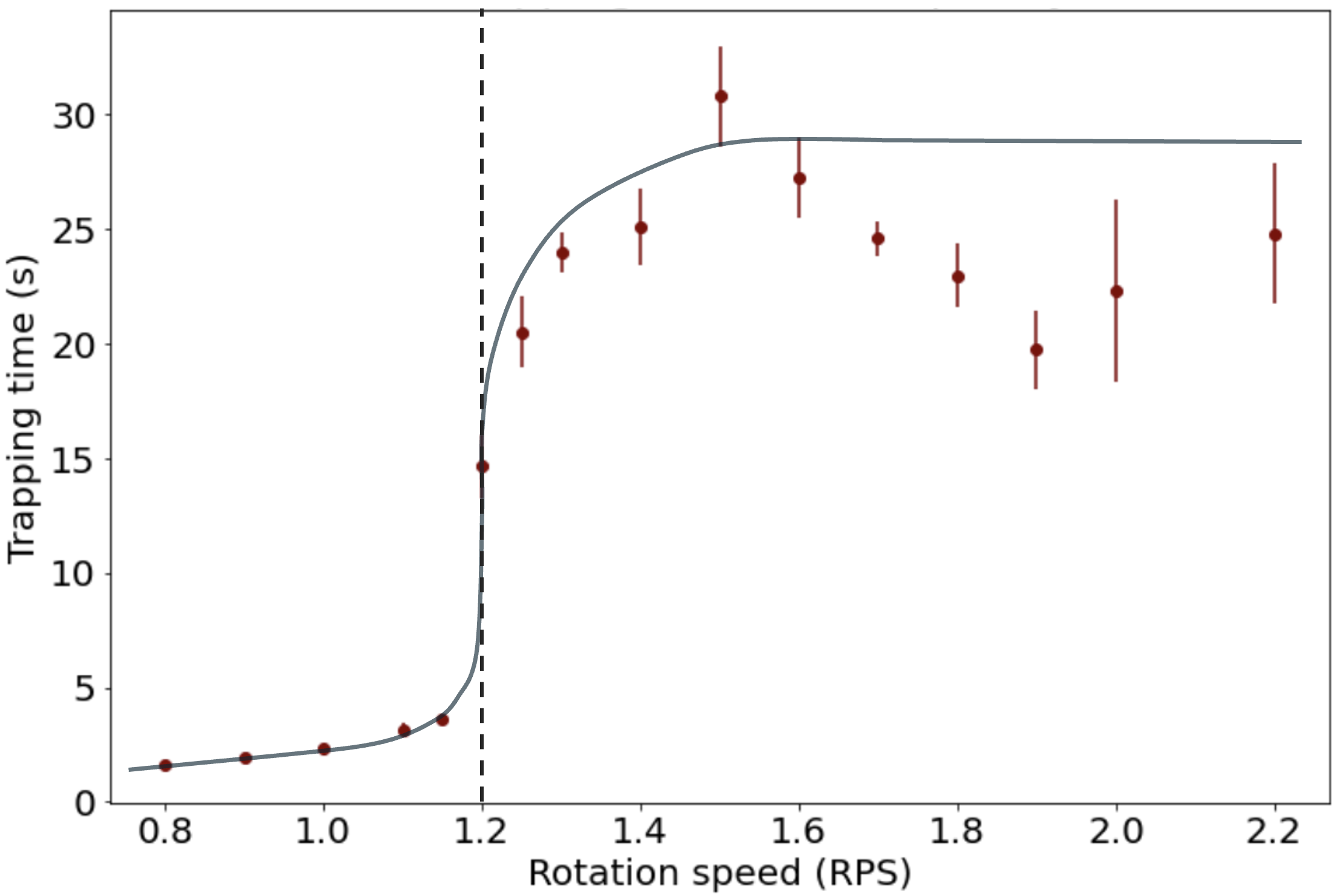}
  \caption{Measured lifetime of LN$_2$ bead in rotating saddle as a function of rotation frequency in rps, showing the sharp threshold at $\Omega \approx 1.2$ rps.}
  \label{fig:P3_Positive_Lifetime_Vs_Frequency_v2}
\end{figure}

After students recognize the instability, increase the rotation frequency incrementally, by approximately 50 millihertz (for instance, from 1.15 to 1.2 rps for the asymmetric saddle). Highlight to students that this extremely small change in rotation frequency, barely noticeable by the naked eye, dramatically alters the system's behavior. Repeat the dispensation of LN$_2$ at the increased frequency. With the asymmetric saddle at 1.2 rps, typical trapping times increase dramatically, shifting from nearly immediate ejection to average lifetimes exceeding 15 seconds. Small differences in initial placement or velocity do not significantly affect trapping success, illustrating the robustness and sensitivity of the system at this threshold. This sharp threshold behavior is clearly illustrated in Fig.~\ref{fig:P3_Positive_Lifetime_Vs_Frequency_v2}, which shows measured trapping times of LN$_2$ droplets as a function of rotation frequency.

A final trial at a higher frequency (approximately 1.5 rps for the asymmetric saddle) demonstrates even more pronounced stability, often yielding trapping times exceeding 30 seconds. For the nearly symmetric saddle ($\beta = 1.06$), the stability region will be shifted slightly but still observable. Instructors should perform similar incremental trials, emphasizing how subtle differences in saddle geometry affect the stability threshold.

\subsection{Mapping stability with dimensionless parameters}
\label{sec:mapping_stability}

\begin{figure*}[!t]
\centering
\includegraphics[width=0.80\textwidth]{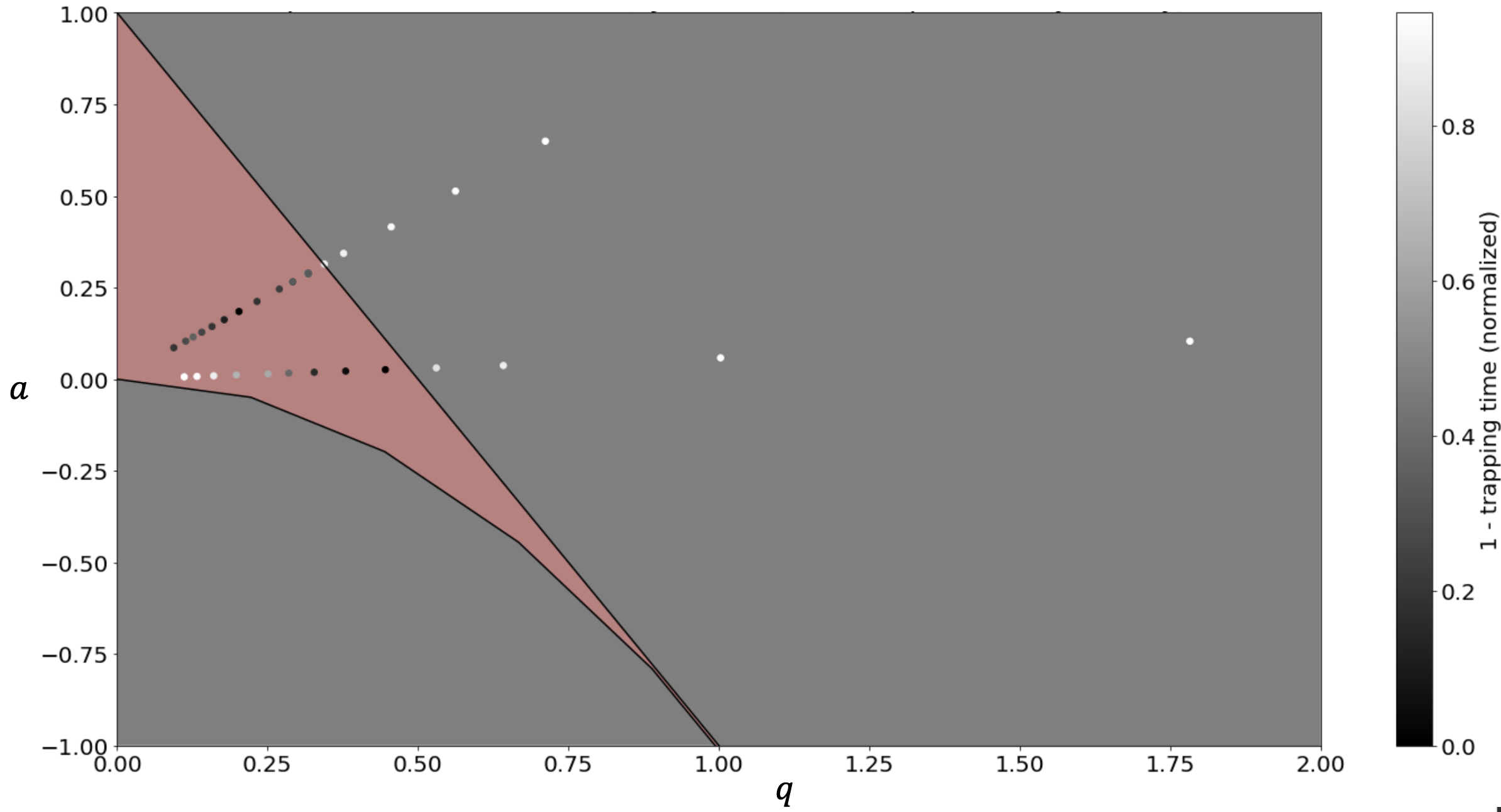}
\caption{Measured stability regions for LN$_2$ droplets in rotating saddles with asymmetry parameters $\beta = 2.67$ (asymmetric) and $\beta = 1.06$ (nearly symmetric), plotted on the $a$–$q$ stability diagram. The grayscale indicates normalized lifetime, where darker points represent longer lifetimes.}
\label{fig:P3_P2_aq_diagram_Final}
\end{figure*}

To visualize the dependence of particle stability on rotation frequency and saddle geometry, we plot the measured lifetimes for LN$_2$ droplets as a function of the dimensionless stability parameters $a$ and $q$ in a composite stability diagram, shown in Fig.~\ref{fig:P3_P2_aq_diagram_Final}. This figure shows data from both the nearly symmetric ($\beta = 1.06$) and asymmetric ($\beta = 2.67$) saddles, allowing direct comparison of their stability boundaries and threshold sharpness. Each data point represents an experimental trial at a given rotation frequency $\Omega$, translated into $a$ and $q$ using Eq.~\eqref{eq:a_q_definition}.

The grayscale intensity of each data point indicates the normalized trapping lifetime: completely white corresponds to the longest recorded lifetime (normalized to 1.0), while completely black indicates immediate bead ejection (zero normalized lifetime). Intermediate shades represent proportionally shorter lifetimes. Thus, this plot visually highlights the sharp threshold behavior characteristic of rotating saddle stability, where even small changes in rotational frequency (and thus small changes in $a$ and $q$) can lead to dramatic differences in particle confinement times. The asymmetric saddle ($\beta = 2.67$) data points near the stability threshold correspond exactly to the sharp transition in trapping lifetime previously illustrated in Fig.~\ref{fig:P3_P2_aq_diagram_Final}, but are now plotted explicitly in terms of the dimensionless stability parameters $a$ and $q$. In contrast, as we will show later, ball bearings exhibit a notably more gradual and less distinct threshold in the $a$–$q$ stability space.

For a given saddle geometry (defined by $\beta$, $h_0$, and $r_0$), varying the rotation frequency $\Omega$ traces out a straight line in the $a$–$q$ stability space, described by:
\begin{equation}
a = 2\frac{\beta - 1}{\beta + 1} q.
\label{eq:a_vs_q_beta}
\end{equation}
This explains why experimental points for each saddle in Fig.~\ref{fig:P3_P2_aq_diagram_Final} lie along distinct linear trajectories.

\subsubsection{Dispensing liquid nitrogen onto the saddle}

Pipettes and other traditional dispensers often become clogged with condensation, making them unreliable for delivering LN$_2$ to a rotating saddle. The most consistent method we found involved an 8-ounce styrofoam cup with a small (approximately 2\,mm-wide) hole punched near the lower rim. Holding the cup above the center of the saddle and gently tipping it allows for controlled pouring through the punched hole.

To maximize observable trapping time, the initial pool of LN$_2$ on the saddle should be roughly the size of a quarter. In stable conditions, most of the liquid near the edges quickly runs off, leaving a few smaller, self-forming droplets near the center. We define the “lifetime” of a dispensation as the duration between the initial pour and the final bead’s departure from the saddle. Fig.~\ref{fig:ln2_evolution} shows the evolution of LN$_2$ droplets after dispensation.

After several trials—or after a particularly long-lived run—condensation may accumulate at the saddle's center. If this occurs, the saddle should be warmed using a hair dryer or heat gun, and any remaining water wiped away. This step is essential to prevent immediate boiling and to maintain the Leidenfrost effect for subsequent trials.

\section{COMPARATIVE RESULTS: LN$_2$ vs. BALL BEARINGS}

While rotating saddles traditionally employ rolling objects such as ball bearings to illustrate dynamic stability, we have demonstrated significant pedagogical advantages in using droplets of liquid nitrogen (LN$_2$). Below we explicitly compare these two approaches, emphasizing differences in fidelity to theoretical predictions, clarity of visual demonstrations, and overall instructional impact.

\subsection{Stability threshold and lifetime comparisons}

\begin{figure*}[!t]
\centering
\includegraphics[width=0.8\textwidth]{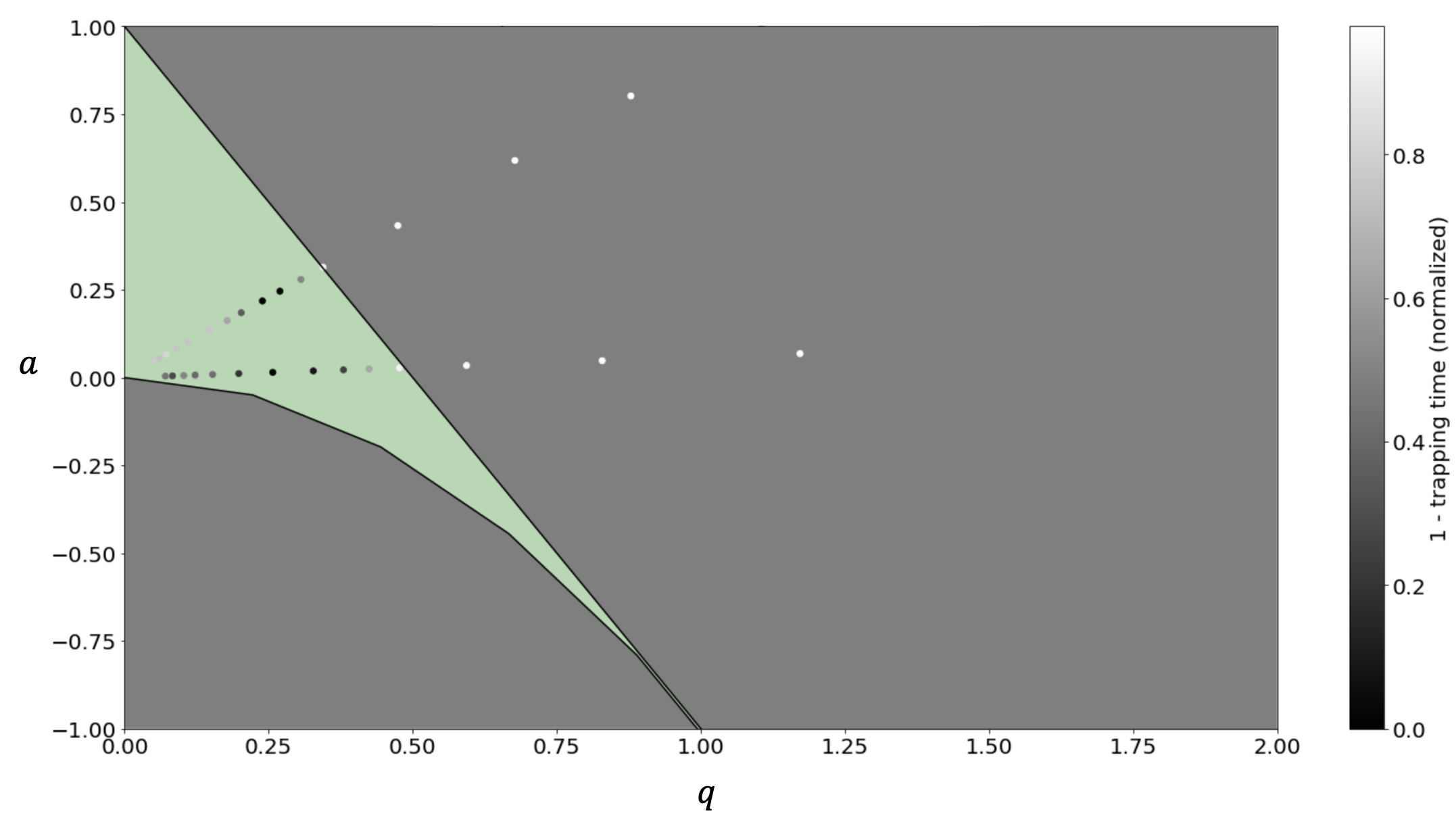}
\caption{Measured stability regions for ball bearings in rotating saddles with asymmetry parameters $\beta = 2.67$ (asymmetric) and $\beta = 1.06$ (nearly symmetric), plotted on the $a$–$q$ stability diagram. The grayscale indicates normalized lifetime, with darker points representing longer lifetimes.}
\label{fig:p3bb}
\end{figure*}

As previously demonstrated, the LN$_2$ bead exhibits a sharply defined stability threshold around a rotation frequency of approximately 1.2 rps (Fig.~\ref{fig:P3_Positive_Lifetime_Vs_Frequency_v2}). This precise threshold results from the minimal friction and negligible internal rotation associated with Leidenfrost levitation, closely matching theoretical predictions for a frictionless particle.

In contrast, ball bearings exhibit a more gradual transition to stability, occurring at slightly higher rotational frequencies (approximately 1.25–1.3 rps in our setup). This less distinct threshold results from frictional effects and internal rolling motion, causing trajectories to deviate notably from theoretical predictions \cite{Thompson_2002, Fan_2017}. Fig.~\ref{fig:p3bb} explicitly illustrates these differences by plotting measured trapping lifetimes for ball bearings as a function of the dimensionless stability parameters $a$ and $q$ for both asymmetric ($\beta = 2.67$) and nearly symmetric ($\beta = 1.06$) saddles. Unlike the sharp, clearly defined threshold observed with LN$_2$ droplets (Fig.~\ref{fig:P3_Positive_Lifetime_Vs_Frequency_v2} and Fig.~\ref{fig:P3_P2_aq_diagram_Final}), ball bearings display a broader stability region and a less distinct boundary. Additionally, the segment of stable trapping (darkest data points) is notably shorter compared to LN$_2$ droplets, further highlighting the diminished instructional clarity inherent to demonstrations relying on rolling objects.

\subsection{Particle trajectory comparisons}

\begin{figure}[h]
    \raggedright
    \begin{minipage}[b]{0.48\textwidth}
        \includegraphics[width=\linewidth]{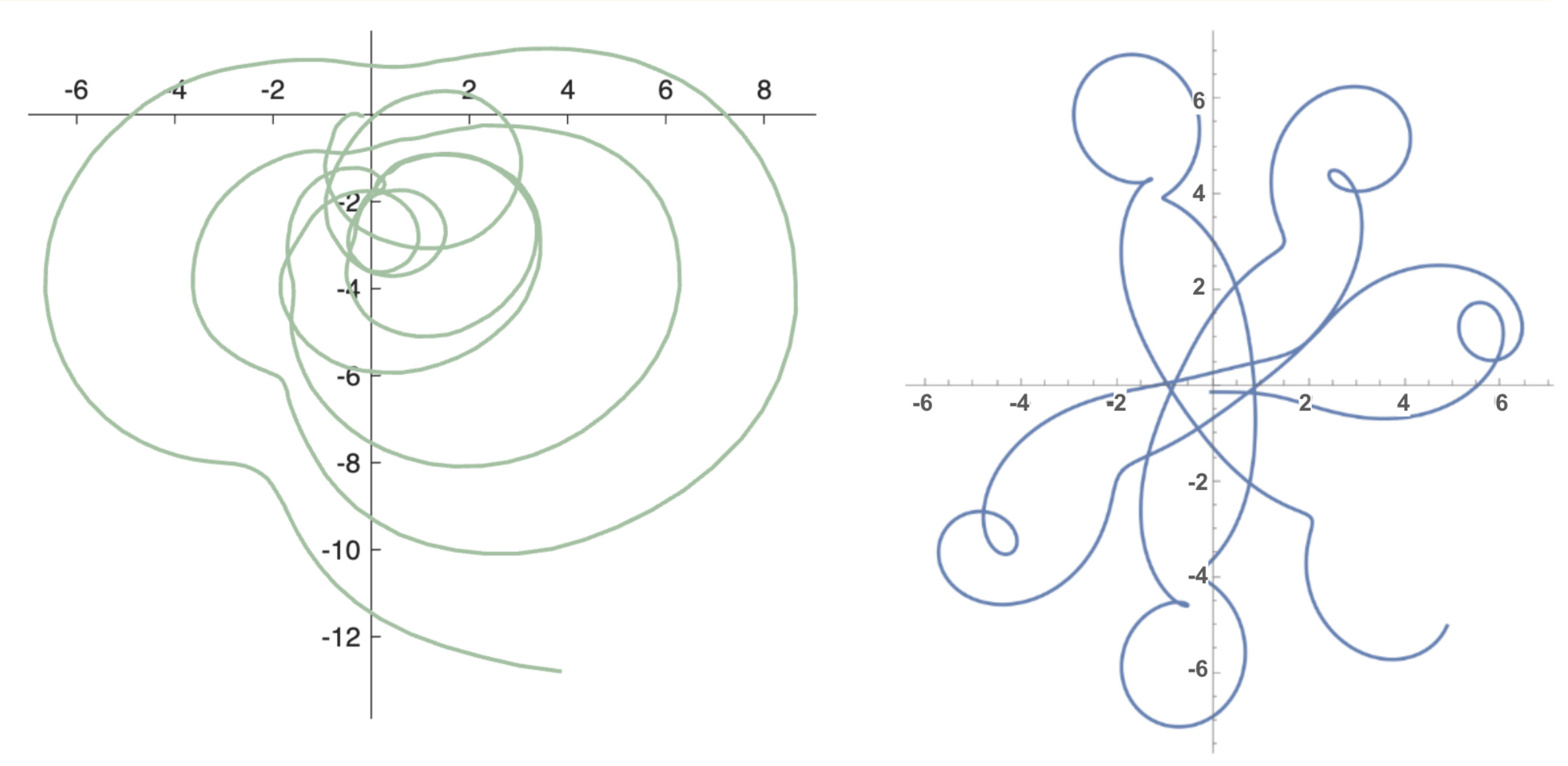}
        \caption{A comparison of the observed trajectory of a ball bearing (left) vs. the predicted trajectory of an ideal particle (right) in a saddle with asymmetry parameter $\beta = 1.06$ and rotational frequency $\Omega = 1.2$ over a period of 8.91 seconds. Initial conditions: $(x_0, y_0, v_{x0}, v_{y0}) = (-0.052~\text{cm}, -0.125~\text{cm}, 0.152~\text{cm/s}, -0.125~\text{cm/s})$. The observed trajectory spirals outward exponentially, exhibiting a smoothed trajectory due to friction and rolling effects.}
        \label{fig:ball}
    \end{minipage}
    \hfill
    \begin{minipage}[b]{0.48\textwidth}
        \includegraphics[width=\linewidth]{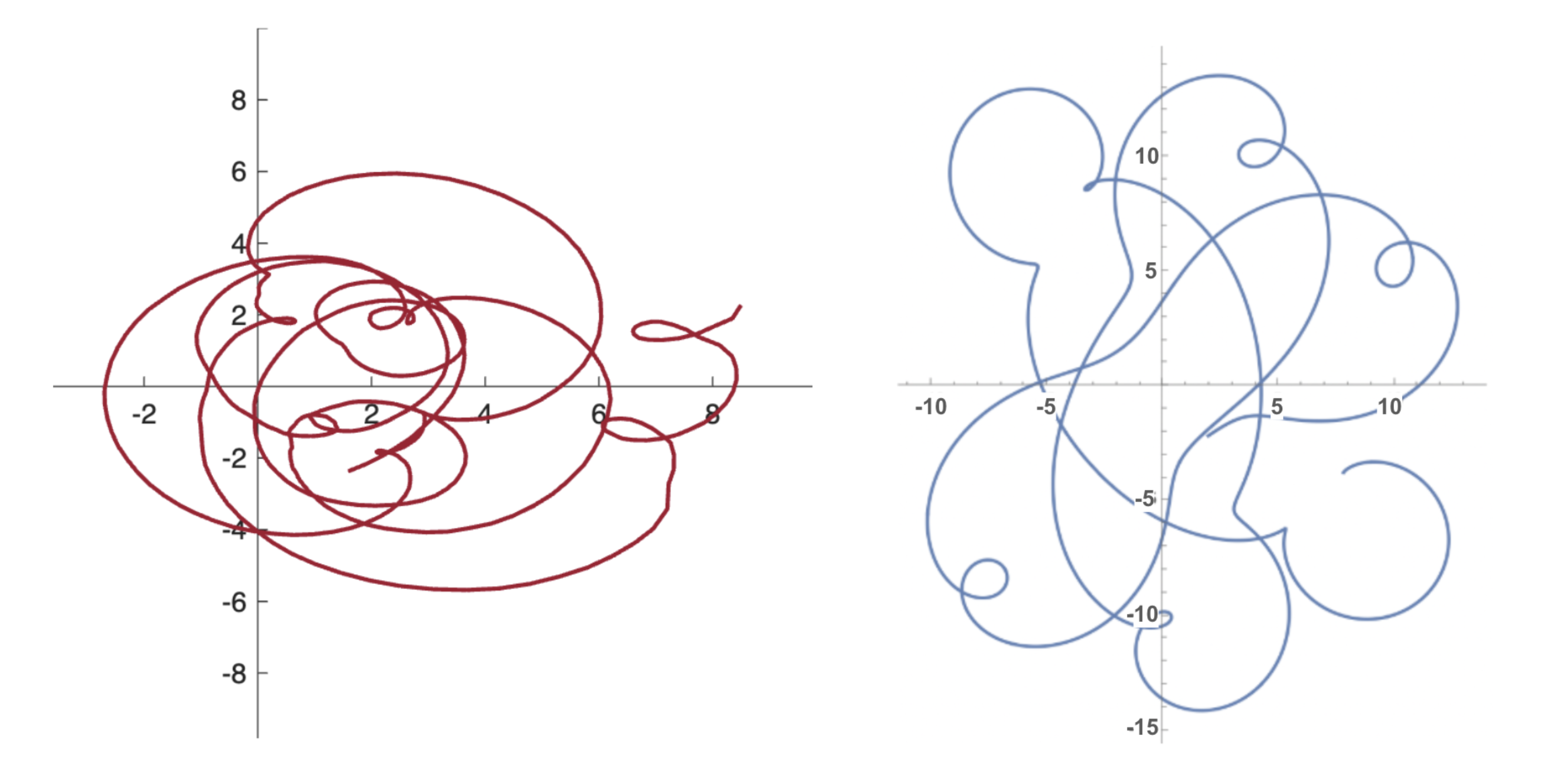}
        \caption{A comparison of the observed trajectory of LN$_2$ (left) vs. the predicted trajectory of an ideal particle (right) in a saddle with asymmetry parameter $\beta = 1.06$ and rotational frequency $\Omega = 1.2$ over a period of 9.28 seconds. Initial conditions: $(x_0, y_0, v_{x0}, v_{y0}) = (1.953~\text{cm}, -2.171~\text{cm}, 1.646~\text{cm/s}, 2.405~\text{cm/s})$. The observed trajectory more closely resembles theoretical predictions, clearly illustrating the rotating ponderomotive-like nature of the confining force.}
        \label{fig:ln2}
    \end{minipage}
\end{figure}

Figures~\ref{fig:ball} and \ref{fig:ln2} directly illustrate the fundamental differences in particle dynamics between ball bearings and LN$_2$ droplets. Fig.~\ref{fig:ball} shows the observed trajectory of a ball bearing (left), characterized by an exponential outward spiral with smoothed features resulting from friction and rolling effects. To an observer, these trajectories appear visibly dominated by surface interaction rather than by the expected potential dynamics.
These non-ideal effects clearly suppress the intricate trajectory structure predicted for an ideal frictionless particle (right), reducing both visual clarity and conceptual coherence.

In sharp contrast, Fig.~\ref{fig:ln2} displays the trajectory of an LN$_2$ droplet (left), which more closely resembles the theoretical prediction (right). The Leidenfrost levitation significantly reduces friction, allowing trajectories that visibly retain the intricate, rotating ponderomotive-like features predicted by theory. An observer can readily discern the complex orbital patterns expected from the theoretical model, making the underlying physics immediately apparent. Consequently, its trajectory clearly illustrates the rotating saddle's confining force, substantially enhancing the demonstration's visual and conceptual impact.

These visual comparisons highlight the pedagogical advantage of LN$_2$ droplets over traditional ball bearings, emphasizing their enhanced fidelity to theoretical dynamics and greater effectiveness as instructional tools.


\section{APPLICATIONS AND EDUCATIONAL VALUE} 

Quadrupole ion traps (RF Paul traps) have broad relevance in modern physics, underpinning essential technologies such as atomic clocks and quantum computing. Incorporating the rotating saddle demonstration into undergraduate courses—such as classical mechanics, electromagnetism, or introductory quantum mechanics—provides tangible, real-world context to otherwise abstract theoretical principles.

Moreover, this demonstration offers an engaging platform for active student participation, prediction, and discussion. Given its simplicity, clarity, and direct analogy to modern experimental systems, it can be effectively adapted into interactive lecture demonstrations, group laboratory exercises, or student-driven final projects.

\section{CONCLUSION}

We have demonstrated significant improvements over the traditional rotating saddle demonstration of dynamic stability by employing liquid nitrogen (LN$_2$) droplets levitating via the Leidenfrost effect. Unlike standard rolling ball bearings, LN$_2$ droplets minimize friction and internal rotation, more accurately approximating the theoretical dynamics of an ideal, frictionless particle. Furthermore, subtle saddle asymmetries provide additional realism, closely modeling potentials encountered in actual RF Paul traps.

These modifications lead to a clearer, more accurate, and more pedagogically relevant demonstration. Specifically, LN$_2$ droplets exhibit a sharper and more distinct stability threshold compared to traditional ball bearings, effectively eliminating common misconceptions caused by frictional and rolling dynamics. We recommend this improved demonstration for clearly illustrating dynamic stability principles in undergraduate physics instruction.

\section*{ACKNOWLEDGMENTS}
We are grateful for helpful discussions with Gregory Kestin and Logan McCarty. We gratefully acknowledge support from the Faculty of Arts and Sciences at Harvard University. GPT-4~\cite{OpenAI2023ChatGPT4} was used to refine this manuscript.

\section*{AUTHOR DECLARATIONS}
The authors have no conflicts to disclose.

\section*{APPENDIX}

\subsection{Saddle fabrication}

Saddles of varying asymmetry were designed using .stl files generated by a Python script (Appendix: Python script for saddle geometry). Saddles were printed using PETG filament on a Prusa Original i3 printer. Models were scaled to 173$\%$ of their original size for an 18 cm square base, printed with 30$\%$ infill for durability. Dark filament colors improved visibility. Print times ranged from 33 to 52 hours, consuming roughly two-thirds of a kilogram per saddle.

Once printed and cooled, saddles were sanded with an electric sander, followed by hand-sanding with fine-grit sandpaper. Care was taken to preserve the intended curvature. Saddles were then coated with Smooth-On XTC-3D under a fume hood, curing overnight. Minor surface imperfections post-curing were gently sanded by hand.

Optionally, saddles were mounted to wooden boards slightly larger than the saddle bases to address any minor deformations. A central hole drilled into the saddle base and board allowed alignment with a dowel, centering the saddle. Hot glue secured the saddle to the board, correcting slight deformations.

A cylindrical motor mount was 3D-printed separately using PLA filament. The mount included openings for the motor axle and centering dowel. Four self-tapping screws attached the mount to the board, with a horizontal screw securing the motor axle.

\subsection{Setup Assembly}

The saddle was rotated using a StepperOnline Nema 17 Bipolar 2A stepper motor controlled by a StepperOnline CNC Motor Driver. A DC power supply providing at least 14.5V was connected to the motor driver. The stepper motor was configured to 3200 steps per full rotation, with frequency controlled by a function generator (frequency monitored via an oscilloscope if necessary).

To minimize friction and improve rotation accuracy, the assembled saddle was elevated using a plywood rectangle attached to the motor. This plywood was secured with screws and wing-nuts to metal rails clamped to the demonstration table.

\subsection{Materials and cost}
The table below summarizes the primary components required for the rotating saddle demonstration, including vendors, estimated costs, and relevant comments. Although the table may appear later due to float placement, it is part of this subsection.

\squeezetable

\begin{table*}[t]
\centering
\squeezetable
\caption{Materials and approximate costs of demonstration setup.}
\begin{ruledtabular}
\begin{tabular}{cllll}
Item & Name & Vendor and Model & Price (USD) & Comment \\ \hline
1 & PETG filament & Prusa Prusament & 29.99 & Price per kilogram. Dark colors recommended for optimal demonstration visibility. \\
2 & 3D-print coating & Smooth-On XTC-3D & 33.81 & \\
3 & Stepper motor & StepperOnline Nema 17 Bipolar 2A (17HS19-2004S1) & 9.62 & Available in bulk at a discount \\
4 & Microstep Driver & StepperOnline CNC Stepper Motor Driver & 28.99 & \\
5 & ¾” wooden board & --- & --- & Optional for use in saddle mount assembly \\
6 & Self-tapping flat-head 8-32 screws & --- & --- & To affix the 3D-printed mount to the saddle \\
7 & 3D printer & Prusa Original i3 MK3S & From 899 & Additional expenses for optional accessories, including textured print sheets \\
\end{tabular}
\end{ruledtabular}
\end{table*}

\subsection{Python script for saddle geometry}
A Python script used to generate .stl files for the saddle geometry is included at the end of the document. Users can modify the script to explore different asymmetries and curvatures.

\begin{figure}[h]
\centering
\includegraphics[width=86mm]{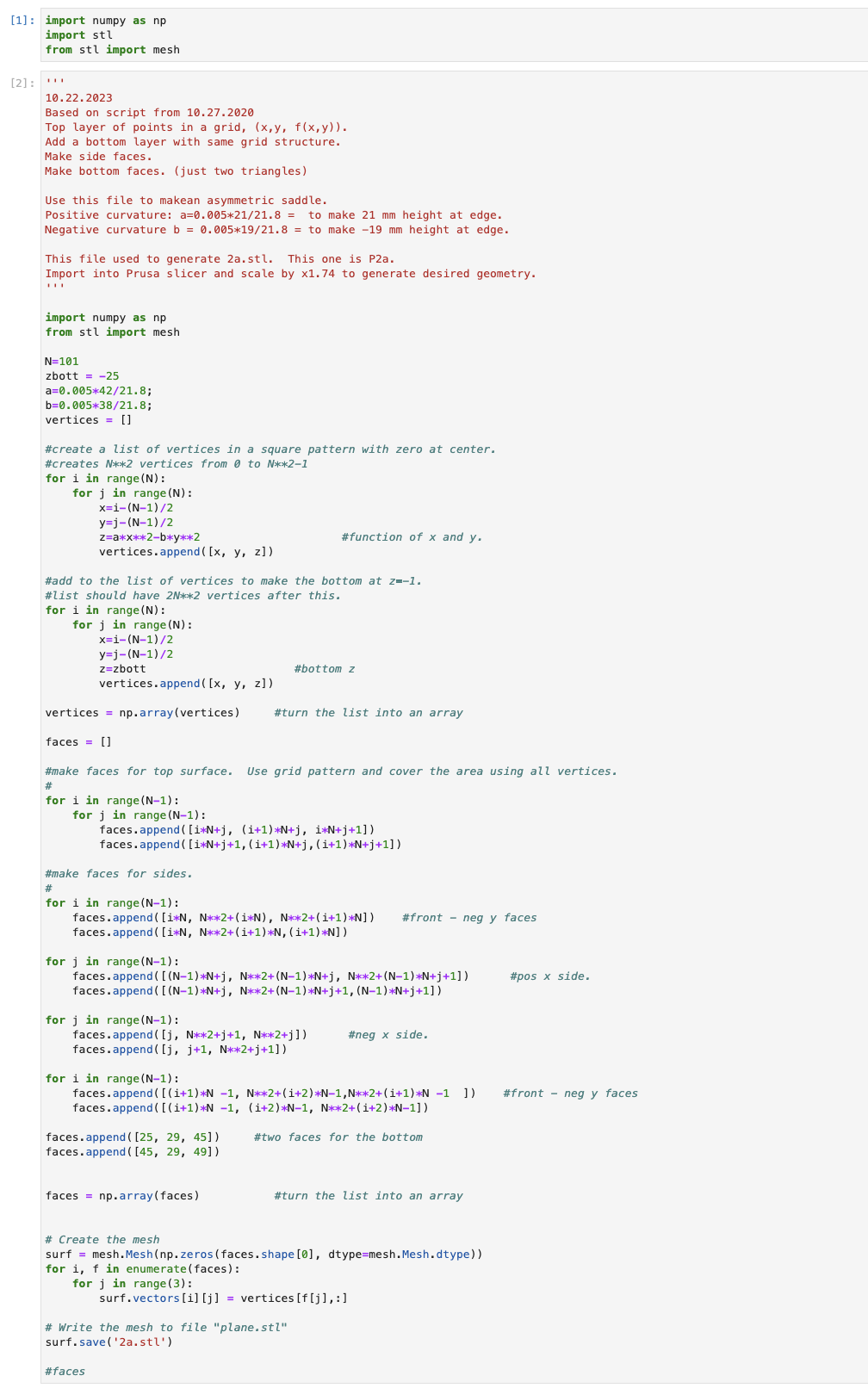}
\caption{Python script for generating customizable saddle geometry. Users may vary parameters to achieve different asymmetries and curvatures.}
\end{figure}

\nocite{*}
\bibliography{references} 


\end{document}